\documentclass[10pt]{article}

\usepackage{graphics}   
\usepackage{endnotes}            
\usepackage{amsfonts}            
\usepackage{amssymb}             
\usepackage{amsthm}              
\usepackage{amsmath}            
\usepackage{algorithm}           
\usepackage{algorithmic}
\usepackage[letterpaper]{geometry}
\usepackage{rotating}            
\usepackage{wrapfig}

\usepackage{setspace}                
\usepackage{natbib}
\bibpunct{(}{)}{;}{a}{,}{;}

\begin{document}

\title{
Transition State Theory for Network Dynamics\thanks{This work was supported in part by NIH award 1R01GM144964-01 and NSF award SES-2448652.  The author thanks Rachel Martin for her helpful input.}
}

\author{
Carter T. Butts\thanks{Departments of Sociology, Statistics, Computer Science, and EECS and Institute for Mathematical Behavioral Sciences; University of California, Irvine; Irvine, CA 92697; \texttt{buttsc@uci.edu}}
}
\date{3/6/2026}
\maketitle

\begin{abstract}
Many classic questions of structural theory concern discrete changes, such as the formation or dissolution of groups, role turnover, or faction realignment.  Here, we consider a basic framework combining prior work on change paths and recent advances in dynamic network modeling with ideas from transition state theory.  This framework facilitates both characterizing the process of structural change and, in some cases, predicting it.  Notably, this approach allows approximate prediction of network change from cross-sectional models, under limited assumptions regarding the underlying microdynamics.   We apply this framework to a simple model of faction realignment in small groups, showing that the process through which realignment occurs can be well-predicted \emph{ex ante} for a number of different network micro-processes.\\[5pt]
\end{abstract}

\theoremstyle{plain}                        
\newtheorem{axiom}{Axiom}
\newtheorem{lemma}{Lemma}
\newtheorem{theorem}{Theorem}
\newtheorem{corollary}{Corollary}

\theoremstyle{definition}                 
\newtheorem{definition}{Definition}
\newtheorem{hypothesis}{Hypothesis}
\newtheorem{conjecture}{Conjecture}
\newtheorem{example}{Example}

\theoremstyle{remark}                    
\newtheorem{remark}{Remark}



The past several decades have seen considerable advance in our ability to model network dynamics, driven largely by empirically calibrated stochastic modeling frameworks including the stochastic actor-oriented models \citep{snijders:sm:2001}, temporal and separable temporal exponential family random graph model (TERGM and STERGM) families \citep{robins.pattison:jms:2001,hanneke.xing:ch:2007,krivitsky.handcock:jrssB:2014}, relational event models \citep{butts:sm:2008}, DyNAMs \citep{stadtfeld.et.al:sm:2017}, and ERGM generating processes (EGPs) \citep{butts:jms:2024}.  These developments have occurred in tandem with the increasing maturity of cross-sectional network modeling schemes, most notably the exponential family random graph modeling (ERGM) framework \citep{robins.et.al:sn:2007,lusher.et.al:bk:2012,schweinberger.et.al:ss:2020}.  These frameworks have made it increasingly possible to specify both \emph{de novo} and empirically calibrated models that capture the mechanisms of structural evolution in a realistic way, and that can recapitulate observed cross-sectional structure in social (and other) networks.  

An emerging challenge in this area is the ability to \emph{summarize} such dynamics in an intuitive and meaningful way, and where possible to \emph{predict} how change might occur when the details of the underlying dynamics are not precisely known (as remains the case in the vast majority of settings, for which the data needed to calibrate most dynamic network models is not available).  This is particularly true for the dynamics that are involved in discrete changes of the type long considered of importance by structural theorists, including the formation and dissolution of groups and larger social aggregates \citep{cohen:bk:1964,tainter:bk:1988,carley:asr:1991} (or the inhibition thereof \citep{simmel:ajs:1898}), the formation and elaboration of hierarchies \citep{blau:asr:1970,gould:ajs:2002}, the displacement and replacement of individuals in social roles \citep{white:bk:1970}, and the realignment of group loyalties \citep{zachary:jar:1977}.  

\begin{figure}
\centering
\includegraphics[width=0.4\textwidth]{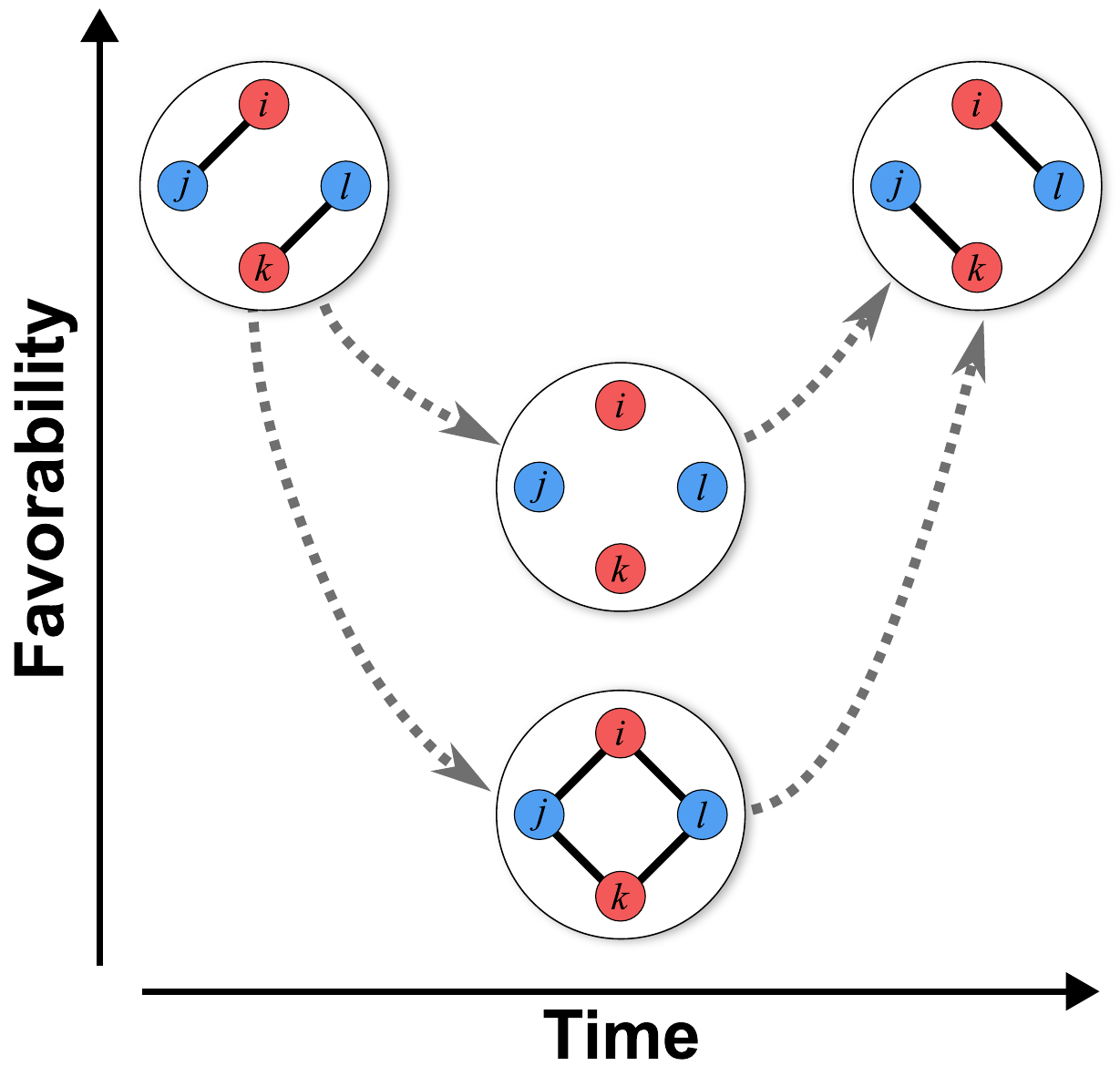}
\caption{Schematic of partner swapping in a heterosexual romantic network (per \citet{bearman.et.al:ajs:2004}).  Alternative pathways require the network to pass through states of differing favorability; those involving highly unfavorable states such as 4-cycles are predicted to be less likely than those involving less unfavorable states (e.g., null graphs). \label{f:swap}}
\end{figure}

Such phenomena also motivate slightly different questions from those that motivate work on quotidian network evolution.  In particular, it is natural to think of these phenomena in terms of changes from one discrete state to another (e.g., a set of unmobilized organizations to an emergent response network \citep[as in][]{butts.et.al:joss:2012}, a single unified group to two independent groups \citep[as in][]{zachary:jar:1977}, or an organization with conserved formal structure but different individuals occupying key roles \citep[as in][]{white:bk:1970}), and to ask \emph{how} such transformations occur.  Intuitively, events involving large-scale structure formation, dissolution, or modification will typically require multiple distinct steps, the nature and order of which are socially consequential.  To take a modest, micro-level example, consider a role change involving a ``partner swap'' involving two men and two women within a heterosexual romantic network \citep[as in, e.g.][]{bearman.et.al:ajs:2004}.  (See Fig.~\ref{f:swap}.)  To effect the change, two new ties within the initially non-partnered cross-sex pairs must be formed, and the two initially existing ties must be broken.  One pathway to such change could entail the sequential breaking of the initial ties, followed by the sequential addition of the new ties.  A very different pathway could entail first the sequential addition of the new partnerships, followed by the severing of the previous ties.  Other possibilities can be enumerated.  While all such pathways will suffice to effect the swap, they have very different implications for the individuals involved (and the properties of the network as a whole).  For instance, the first pathway requires the four-person subgraph to pass through a period as an empty graph, with (discounting the possibility of outside ties) all members forgoing romantic partnerships.  On the other hand, the second pathway requires the formation of a four-cycle, with all members not only having concurrent partnerships, but forming them with their partners' partners.  As \citet{bearman.et.al:ajs:2004} observe in their study of adolescent romantic relationships, this seems to be highly disfavored (at least within their study population), suggesting that this pathway is likely to be blocked in practice.  All change paths are not, therefore, equally likely - and, in particular, paths that require the system to transit through highly unfavorable states will tend, \emph{ceteris paribus,} to be less likely than those transiting favorable states.  \citet{butts.carley:jms:2007} argue for such path selection preferences in networks controlled by individuals or groups (e.g., formal organizational or infrastructure networks), but as Bearman et al.'s more romantic example illustrates, similar effects are also expected to arise in decentralized settings.  This point was also made by \citet{willer:jms:2007}, who observed that biased path selection will arise in networks whose evolution is controlled by individual actors with preferences over network states.  Of course, neither micro-level nor macro-level decision making processes are necessary for this effect to obtain: whatever the underlying mechanism, so long as the microdynamics of the evolving network favor some states over others, path selection effects will result.

The impact of favorable or unfavorable states on change paths has other implications as well.  For instance, changes that \emph{require} the system to pass through highly unfavorable states may be predicted to have slower dynamics than those that do not.  More subtly, change paths that involve moving through unfavorable states via a series of  ``islands'' of favorable states run the risk of becoming ``stuck'' in one or more of these intermediate waypoints, delaying or possibly reversing progress towards the target state.  In an intervention context, this may suggest options for accelerating or inhibiting change by altering the favorability of the intermediates (or of the unfavorable states through which the system must transition). These types of path selection effects have been independently noted in many different fields, being perhaps most deeply considered in the well-known transition state theory of reaction kinetics \citep[see e.g.][]{dill.bromberg:bk:2010}.  Here, we will make use of some of these ideas, although we will develop them within the context of network dynamics.

In what follows, we propose some first elements of a transition state theory for network dynamics.  As noted, we build on a number of existing elements both within and beyond the network field, most notably the ideas of change paths (and their differential favorability) advanced by e.g. \citet{butts.carley:jms:2007} and \citet{willer:jms:2007}, work on graph potentials and related dynamic processes (particularly EGPs, with key contributions for this work including \citet{snijders:sm:2001}, \citet{koskinen.snijders:jspi:2007}, and \citet{butts:jms:2024}), and transition state theory within the molecular sciences \citep{laidler.king:jpc:1983}.  As we will observe from the outset, the theory does not have unlimited scope, instead being intended to provide useful approximations to and predictions for the behavior of networks that evolve along a potential surface, and whose dynamics tend to favor uphill moves.\footnote{Throughout, we follow the social science sign convention of high-probability states as being of high potential, rather than the physical science convention of high-probability states as being of low potential.  Nor does the graph potential refer to a potential energy surface, though it is related to the free energy.}  These conditions, as we will show, describe many models currently used in the social network literature.  Where applicable, the theory can be quite powerful, allowing qualitative (and potentially quantitative) predictions for network change that can be made using only cross-sectional models, and that are agnostic to micro-dynamic details.  This renders the approach potentially useful in many settings for which cross-sectional network data is available, but dynamics can be constrained only qualitatively (drawing e.g. on ethnographic, observational, or other evidence).

The remainder of the paper is structured as follows.  Sec.~\ref{sec:theory} lays out the framework, including core concepts and some basic computational strategies.  Sec.~\ref{sec:demo} provides an illustrative application of the theory to a simple model of faction realignment, showing how the approach can be used to predict the process by which realignment along competing bases of solidarity would be expected to occur.  Subsection~\ref{sec:egp_compare} puts the predictions of the theory to test against four different models of network microdynamics, showing that - despite their differences - many aspects of their change paths are well-described by the theory.  Sec.~\ref{sec:discussion} considers some additional issues and extensions of the work, and Sec.~\ref{sec:conclusion} concludes the paper.

\section{A Basic Theory of Network Change} \label{sec:theory}

We begin by laying out our basic theory of network change, followed by some remarks on computational strategies for implementation.  These ideas are subsequently applied in Sec.~\ref{sec:demo} to the analysis of a model of faction realignment.  Our framework builds most proximately on established concepts in the literature on ERGMs \citep{lusher.et.al:bk:2012,schweinberger.et.al:ss:2020} and EGPs \citep{butts:jms:2024,snijders:sm:2001,koskinen.snijders:jspi:2007}, and on the change path concepts articulated in the network literature by \citet{willer:jms:2007} and \citet{butts.carley:jms:2007}.  (Compare also to the closely related notion of local graph stability introduced by \citet{yu.et.al:siam:2021}.)  We borrow the notions of transition states and intermediates from chemical kinetics \citep{dill.bromberg:bk:2010}, along with the notion of systematically quantifying progress along a change coordinate, and the idea that local (quasi)equilibrium behavior can be predictive of dynamics.  Although the theory can be applied to models that are not written in ERGM form, we will throughout use ERGM/EGP concepts for ease of exposition and familiarity to a network audience; since any model within the scope of the present development can in principle be written in ERGM form (albeit perhaps not parsimoniously), this is without loss of generality.

\subsection{Concepts}

Consider a random graph $Y$ on finite support $\mathcal{Y}$.  Following \citet{butts:jms:2024}, we associate $Y$ with a \emph{potential}, $q : \mathcal{Y} \mapsto \mathbb{R}$, such that $\Pr(Y=y) \propto \exp(q(y))$. We partition the elements of $\mathcal{Y}$ into a set of disjoint sets, $S$, such that for all $s\in S$, $q(y)=q(y')$ for all $y,y'\in s$.  We refer to $S$ as an \emph{ensemble}, and the elements of $s$ as \emph{states}.  The \emph{state probability} of $s\in S$ is defined as $\Pr(Y \in s)$.  By construction, it is clear that $\Pr(Y\in s) \propto \exp(q_s) |s|$, where $q_s=q(y)$ for any $y\in s$ and $|\cdot|$ denotes cardinality.

We will be interested in settings in which $Y$ is governed by a dynamic process, $D$, that generates sequences of realizations $(\ldots,y^{(i)},\ldots)$ such that, for all $i$, $y^{(i)}\in \mathcal{Y}$ and $\Pr(Y^{(i)}=y^{(i)}) \propto \exp\left(q\left(y^{(i)}\right)\right)$.  We refer to any sequential pair $(Y^{(i)},Y^{(i+1)})$ arising from $D$ as a \emph{transition}.  A specific transition from graph $y$ to graph $y'$ is said to be an \emph{allowable transition} for $D$ iff, for all $i$, $\Pr(Y^{(i)}=y,Y^{(i)+1}=y)>0$.  Similarly, we say that there is an allowable transition from state $s$ to $s'$ iff there exist $y\in s$, $y'\in s'$ such that $y,y'$ is an allowable transition.  Given the above, we may endow $S$ with a \emph{transition structure,} $T_D$, which may be represented by a digraph $T_D=(S,E)$ on the set of states such that $(s,s')\in E$ iff $(s,s')$ is an allowable transition for $D$.  Note that we do not here make any particular assumption regarding the timing of transitions (other than the requirement that inter-transition times be almost surely finite), nor whether the transitions represent a discrete-time process versus discrete transitions in continuous time.  Although we will demonstrate our framework in a continuous-time system (see Sec.~\ref{sec:demo}), it is equally applicable in both regimes. 


With these preliminaries, we may now define the key ideas of our framework.  Following \citet{butts.carley:jms:2007} and \citet{willer:jms:2007}, we define a \emph{change path} on $S$ as a non-repeating sequence of serially adjacent states in $T_D$; i.e., $P=(s,\ldots,s')$ is a change path if $P$ is the vertex sequence of a directed path in $T_D$.  For source $s$ and destination $s'$, there are in general many change paths.  One of which we will make heavy use here is what will call a \emph{maximum state probability change path} (MSPCP).  Let $\mathcal{P}$ be the set of all $s,s'$ change paths on $T_D$.  Then $P^*$ is an MSPCP from $s$ to $s'$ if it is a solution of
\begin{equation}
P^*= \arg\max_{P\in \mathcal{P}} \prod_{s_j \in P} \Pr(Y \in S_j).  \label{e:MSPCP}
\end{equation}

One simple motivation for the MSPCP is as follows.  Imagine that we observe our system at a random time to be in state $s$, and then observe it at a series of random (i.e., well-separated) times thereafter.\footnote{Since our assumptions implicitly require $D$ to be ergodic, we may define some $t$ such that if $Y^{(t)}$ is distributed proportional to $\exp(q(Y))$, $Y^{(t+\Delta)}$ has the same distribution for any $\Delta$.  These may be referred to as ``random times'' vis a vis $D$.}  Given that our observations fall on a $s,s'$ change path in $S$, the most likely path to be observed is a MSPCP.  More intuitively, like a low pass through a mountain range, a MSPCP constitutes a sequence of adjacent states that connect the two endpoints while being highly accessible.  We may thus hypothesize that the MSPCP from $s$ to $s'$ will predict the path taken in an actual transition, given that the transition occurs.

We note immediately that this prediction is heuristic, and will not hold for all systems.  However, typical choices of $D$ have characteristics that favor MSPCP-like change paths.  In particular, most current models of network dynamics propose movement through $\mathcal{Y}$ that is biased ``uphill'' with respect to $q(Y)$, encouraging dynamics to follow sequences of high-probability graphs.  These include utility-theoretic models such as SAOMs \citep{snijders:sm:2001} and DyNAMs \citep{stadtfeld.et.al:sm:2017}, potential-following models such as the LERGM \citep{koskinen.snijders:jspi:2007} and change inhibition (CI) process \citep{butts:jms:2024}, and separable temporal ERGMs \citep{krivitsky.handcock:jrssB:2014} (and their continuum variants \citep{butts:jms:2024}).  The discrete (quasi) time Metropolis and Gibbs Markov chain Monte Carlo (MCMC) algorithms typically used to simulate ERGM realizations \citep{hunter.et.al:jcgs:2012} are also of this type.  A rare counterexample is the differential stability (DS) process \citep{butts:jms:2024}, in which edge changes occur uniformly at random (but change times depend on the potential of the current state).  Thus, while one can identify models for which the MSPCP would not be expected to be a reasonable approximation, these are not typical of those used in current practice.

Another important motivation of the MSPCP is pragmatic: it may be computed from the cross-sectional distribution of $Y$ and the allowable transitions of $D$, without having to specify $D$ in detail.  Since most currently-used network models (including the examples noted above) all share the property that the allowable transitions of $D$ are defined by single edge changes (i.e., Hamming steps), it follows that all of them are covered by the same specification of $T_D$.  Likewise, cross-sectional distributions are available immediately for some classes of processes (notably ERGM generating processes \citep{butts:jms:2024}, which are directly specified in terms of $q(Y)$), and may be approximated by simulation for others (e.g., SAOMs).  Perhaps more interestingly, the MSPCP provides a potential tool for gaining insight into plausible network dynamics in settings for which \emph{only} cross-sectional models are available (e.g., typical ERGM settings).  While it must be emphasized that cross-sectional distributions do not uniquely determine dynamics, \emph{combining} distributional information with simple assumptions about dynamics obtained via other means can provide useful indications of how change would be expected to unfold.  In Sec.~\ref{sec:demo}, we illustrate such an analysis for a simple model of faction realignment.

Finally, we also note that some systems have have high-probability state paths that, while less favorable than the MSPCP, may be favorable enough to be selected a non-vanishing fraction of the time.  Most obviously, there will in practice be a large ``bundle'' of paths that are extremely close to the MSPCP in state space, and that collectively have high probability; these are generally of technical rather than substantive interest, as they by definition represent very similar states to those directly on the MSPCP.  More interesting are alternative pathways that are are not close to the MSPCP, but that are still favorable enough to be observed.  Sec.~\ref{sec:demo} shows an example of such an alternative change path.  We comment further on these secondary paths, and on their identification, in Sec.~\ref{sec:discussion}.

\subsection{Annotating Features of Change Paths}

Examination of the MSPCP provides a number of insights into the plausible pathways through which a specified change may occur.  Many of these are revealed via properties of the states through which the path wends.  It is useful, in summarizing these, to simplify the change path by parameterizing it in terms of a \emph{change coordinate} that varies from a value of 0 at the source to 1 at the destination; in particular, for MSPCP $P^*$, we define the change coordinate at the $i$th state to be $(i-1)/(|P^*|-1)$.  Graph statistics of $s_i \in P^{*}$ may then be plotted against the respective change coordinate to observe how they vary over the course of the process.  (Examples of this are shown in Sec.~\ref{sec:demo}.)

\begin{figure}
\centering
\includegraphics[width=0.7\textwidth]{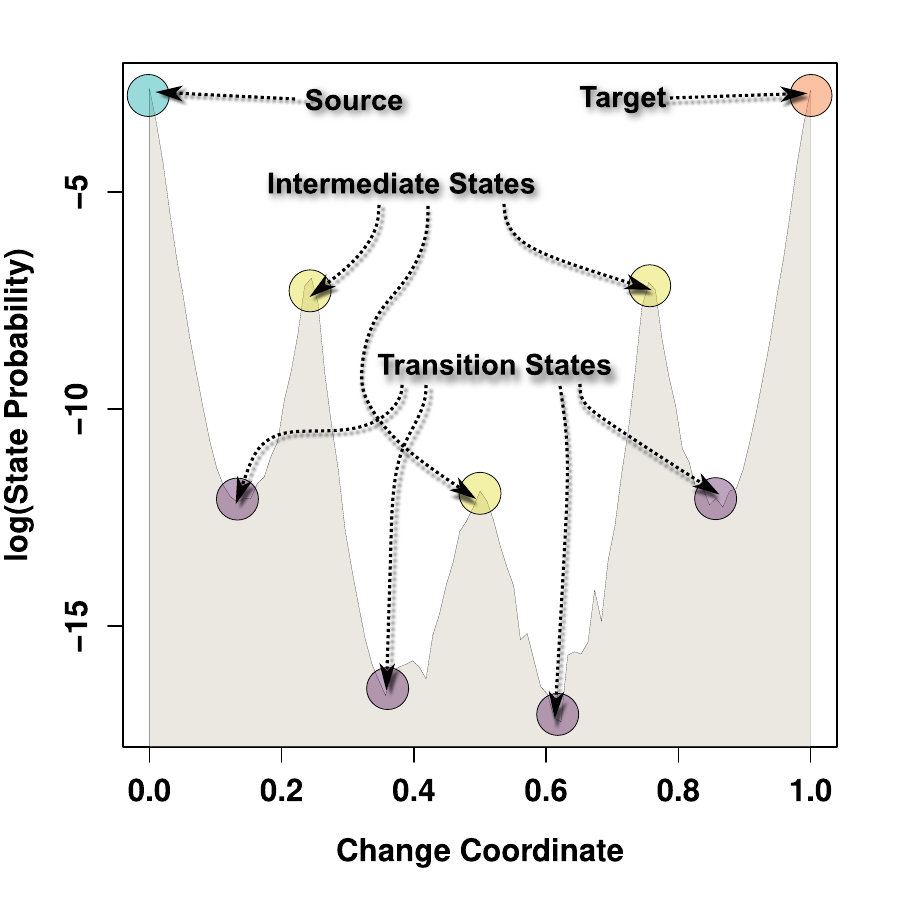}
\caption{Example of a change path, with annotated intermediate and transition states.  The transformation from the source to the target state occurs along the change coordinate; interstitial states of locally maximal probability are known as intermediate states, while local probability minima are transition states.  Deep wells between intermediates can act as barriers to progress along the change path. \label{f:annot}}
\end{figure}

Within a change path, certain states serve as particularly important ``milestones.'' Of special importance are those states whose probabilities are local maxima or minima along the change coordinate.  Borrowing terminology from chemical transition state theory, we refer to the former as \emph{intermediate states,} and the latter as \emph{transition states}.  An example of a change path showing such states is shown in Fig.~\ref{f:annot}. Intermediate states represent states that are locally stable, and hence predicted to be relatively persistent; an evolving network risks becoming ``trapped'' in an intermediate state for an extended period of time, and may require specific changes to advance along the change path.  By turns, transition states represent highly unfavorable states that are rarely occupied for long periods.  For processes that tend to evolve by moving uphill on the probability surface, these states represent ``moats'' of low probability that must be crossed to proceed.  An evolving network has a high risk of becoming ``stuck'' on the near side of such states, and potentially diffusing backwards along the change path before crossing the transition state.  On the other hand, a network that crosses such a barrier will likewise have a low propensity to return, making transition states important markers of persistent change.  Both intermediate and transition states may be identified by plotting the log state probability along the change path.  Where this is computationally infeasible, plotting the potential ($q_s$) along the change coordinate is often a reasonable approximation; the two will correlate strongly where state multiplicities are relatively similar along the change coordinate (since $q_s|s|$ is, up to a constant, the log state probability), but disparities may occur when some states contain many more graphs than others. 

As the above discussion implicitly observes, we do not expect that real trajectories will simply march along the change path: the system may repeatedly backtrack and advance with respect to the change coordinate, with occasional off-pathway ``excursions'' that then return to it.  More precisely, the change path will be embedded within a longer trajectory (a \emph{change walk,} per \citet{butts.carley:jms:2007}).  Following the analogy of a mountain pass, change path predictors such as the MSPCP identify trails that are predicted to be favorable for the intended class of systems, and along which we expect to find the system much of the time; however, our hypothetical hiker may walk too and fro along the trail, wander off the path to examine some interesting landmark, or simply rest for a time on a suitably accommodating log.  Features such as intermediate and transition states thus give insight into aspects of the plausible dynamics beyond the change path itself.  As a change walk could potentially contain more than one change path, we refer here to the most favorable (highest state probability) path within the walk as the underlying change path unless otherwise indicated.

\subsection{Computation}

As noted, a virtue of MSP change paths is that they can be approximated from purely distributional models, as well as from models with explicit dynamics.  A simple approach is outlined as follows.  First, we draw a sample from $Y|q$ using MCMC or related methods starting with a set of overdispersed seeds including the desired source and destination (ensuring coverage of the relevant portions of $\mathcal{Y}$).  Next, we collapse the sampled graphs into states (for ERGMs or EGPs, this may be done using the sufficient statistics); the state probabilities can then be estimated directly from the observed state frequency distribution.  Provided the sampling method has the same allowable transition rule as the target $D$ (as is e.g. the case for most EGPs and SAOMs with edge-toggle MCMC routines) $T_D$ may then be approximated by treating each observed $s,s'$ pair as having an allowable transition if at least one such transition is observed in the sampled trajectories.  An alternative approach is to launch multiple short trajectories following the transition rules of $D$ from graphs sampled from each state, using all observed transitions as an approximation of the edges of $T_D$.  Given the resulting transition graph, weighting each $(s_i,s_j)$ edge by $\log \Pr(Y \in s)$ and seeking a maximum-weight path from the source to the target state (using e.g. Dijkstra's algorithm) will then yield the approximate MSPCP.  Although the full state space is too large to be enumerated for all but the smallest graphs, it should be noted that low probability states are unlikely to appear in the MSPCP, particularly if they are far from the source and destination states, and full enumeration is thus unnecessary.  We employ this strategy in the example below.

Another computational strategy, adapted from an elegant scheme proposed by \citet{koskinen.lomi:jsp:2013} in the context of likelihood computation for LERGMs, is to carry out path sampling by selecting graph pairs $y,y'$ from $s,s'$, and sampling over the space of Hamming trajectories from $y$ to $y'$.  This exploits the fact that any such trajectory can be written as a sequence of edge state ``toggles'' consisting of (1) the set of edge differences between $y$ and $y'$ together with (2) some number of toggle \emph{pairs} involving the same edges.  Applying such toggles in any order will transform $y$ into $y'$, and one can sample from the space of $y,y'$ change walks via MCMC on the space of pairs.  This approach is particularly useful for large or recalcitrant systems in which adequate sampling of the state space is difficult, notably including cases in which the source and destination states are both of low probability.  Note, however, that the path sampling must be replicated over a sample of graphs from both $s$ and $s'$, and additional sampling is required to estimate the state probabilities themselves.  Additionally, MCMC moves that elongate or truncate the trajectory require use of a term that includes the normalizing factor of the graph distribution.  Using the observed states from the sampled change walks as seeds for a larger sample from $Y|q$ (per above) is a reasonable approach to the former problem that ensures coverage of the relevant portion of the state space.  While the latter problem is difficult in general, in the case of ERGMs it can often be solved by using the bridge sampling methods frequently used for ERGM log-likelihood calculation \citep{krivitsky.et.al:jss:2023}.

\section{Application to a Faction Realignment Model} \label{sec:demo}


Subgroups frequently coalesce around shared characteristics or interests, which can serve as bases of solidarity during times of competition or internal conflict \citep{durkheim:bk:1893,pfeffer.salancik:bk:1978}.  Where the pressure to organize into factions (e.g., due to resource competition) occurs in a setting with multiple, poorly correlated attributes, factions may form along some attributes while cutting across others.  During periods of disruption, faction realignment may occur, with a different set of attributes acting as the focal points for division.  The mechanics of such realignments, however, are nontrivial: realignment necessarily requires individuals to break off ties with many of their current allies, while forming affiliations with those to whom they are currently opposed.  Particularly given that such changes cannot happen all at once, it is not obvious how they may come about.  Is realignment more readily triggered e.g. by the initial formation of ``treasonous," cross-faction ties that then reduce the need for (and thus undermine) ties within the existing factions?  Or, alternatively, is realignment more easily triggered by the erosion of internal solidarity (due to a loss of intra-faction ties) which, in turn, motivates members of existing factions to seek support from the ``other side?'' Here, we consider this question within the context of a simple model of faction realignment, using the transition state framework.

\subsection{A Simple Model with Competing Solidarities} \label{sec:demo_mod}


The model we shall employ here presumes a fixed set of individuals, who are connected within an affiliation network; we interpret ties to represent relationships of mutual solidarity or support.  Affiliation is presumed to be costly (as it may require provision of services or resources, taking on risks for the other party, or other difficulties); moreover, as each existing affiliation represents commitments that may restrict ego's freedom of action \citep[per][]{krackhardt:rso:1999}, we assume that the cost of a new affiliation increases with the number of relationships ego already has.  On the other hand, certain circumstances render affiliation more rewarding.  Affiliations are more rewarding with those with whom ego is aligned on certain specific attributes (``bases of solidarity''), and the gains to affiliation are amplified within basis-aligned coalitions (e.g., due to enhanced ability to jointly mobilize to obtain rewards).  Specifically, we assume that each minimal clique involving basis-aligned alters to which ego belongs contributes to his or her payoffs (thus making affiliations to basis-aligned alters more valuable when ego  and alter have multiple basis-aligned shared partners).  Affiliation formation is thus governed by the balance between the cost of sustaining ties of mutual support and the gains to be had by forming coalitions aligned along a basis of solidarity.

We specify a model for the above in ERGM form \citep{schweinberger.et.al:ss:2020}.  We take $N=20$ (in keeping with a small-group setting, in which size does not intrinsically limit affiliation potential or mutual observability), and assume that each individual possesses two dichotomous attributes ($B^1$ and $B^2$) representing potential bases of solidarity.  As our interest is in competing bases of solidarity, we choose a maximally orthogonal scenario: on each basis, 10 individuals are assigned to a value of 0 (otherwise 1) on each attribute, and $B^1$ is uncorrelated with $B^2$.  As we are modeling an unconstrained, small-group setting, we specify our model with respect to the counting measure.  Our ERGM potential is given by
\begin{align}
q(Y) &= \theta^\intercal t(Y)\\
     &= \theta_e t_e(Y) + \theta_{2s} t_{2s} (Y) + \theta_{m1} t_{m}(Y,B^1) + \theta_{m2} t_{m}(Y,B^2) + \theta_{\Delta 1} t_{\Delta}(Y,B^1) + \theta_{\Delta 2} t_{\Delta}(Y,B^2), \label{eq:model}
\end{align}
where $t_e$ is the edge statistic, $t_{2s}$ is the 2-star statistic, $t_m$ is the nodematch statistic, and $t_{\Delta}$ is the local triangle count statistic, i.e.:
\begin{gather}
t_e(Y) = \sum_{i=1}^N \sum_{j=i+1}^N Y_{ij}\\
t_{2s}(Y) = \sum_{i=1}^N \sum_{j=1}^N \sum_{k=j+1}^N Y_{ij} Y_{ik}\\
t_m(Y,Z) = \sum_{i=1}^N \sum_{j=i+1}^N I(Z_i=Z_j) Y_{ij}\\
t_\Delta(Y,Z = \sum_{i=1}^N \sum_{j=i+1}^N \sum_{k=j+1}^N I(Z_i=Z_j) I(Z_j=Z_k) I(Z_i=Z_k) Y_{ij} Y_{jk} Y_{ik}.
\end{gather}

\begin{figure}
\centering
\includegraphics[width=0.6\textwidth]{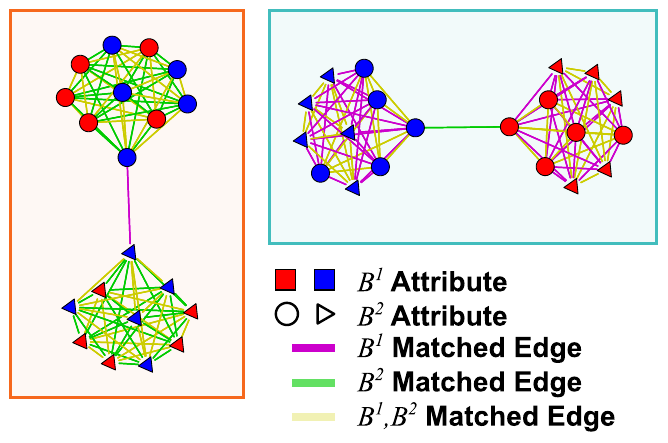}
\caption{Basis-aligned alliance networks for the faction model. Factions may align along attribute $B^1$ (color) or $B^2$ (shape); affiliation costs frustrate simultaneous alignment along both bases of solidarity.  \label{f:demo_aligned}}
\end{figure}

Our parameters are given by $\theta_e=-6$, $\theta_{2s}=0.1$, $\theta_m1=\theta_m2=4$, and $\theta_{\Delta 1}=\theta_{\Delta 2}=1$.  We note in passing that, while triangle terms are avoided in many settings due to their tendency to produce runaway clique formation, in this case the effect is precisely what we are interested in: basis-aligned coalitions become increasingly stable with size, and the network is driven towards forming maximal coalitions along each potential basis of solidarity.  However, this tendency is frustrated by the costs of affiliation, which render a single ``coalition of the whole'' infeasible.  Thus, the network will generally fall into one of two classes of states: two dense factions coalesced around $B^1$ (with a much smaller number of cross-cutting ties between some individuals who align on $B^2$); or two dense factions coalesced around $B^2$ (with, as before, some cross-cutting ties between individuals who align on $B^2$).   High probability realizations of these two states are shown in Fig.~\ref{f:demo_aligned}.

\subsection{State Probabilities and the MSP Change Path}

As specified, our model predicts that, if observed at a random time, our system would be likely to be observed to be split into two factions, aligned either with basis $B^1$ or $B^2$.  We have, however, motivated it at least tacitly from an implicit dynamic foundation: we take $q$ to reflect, in one way or another, the behaviors of our hypothetical actors, whose interactions construct the network.  As such, we may expect that, were our system observed over an extended period, we would eventually observe a realignment in which the relevant basis of solidarity would switch from one attribute to the other.  Our model, as defined, does not allow us to make any such statements.  However, if we further posit that the network evolves by a some dynamic process that is consistent with the specified cross-sectional distribution, then it becomes possible to ask how such realignments occur.  While any EGP would suffice in this regard, there are many candidates, and we may be considerably less confident which of them should be chosen than we are in the overall distributional behavior of the system (or the general factors that should shape it).  Can we nevertheless make some statements that, while heuristic, give insight into how a \emph{range} of such generative processes might behave?  The answer is affirmative, as we can see via application of the maximum state probability change path.

\begin{figure}
\centering
\includegraphics[width=0.8\textwidth]{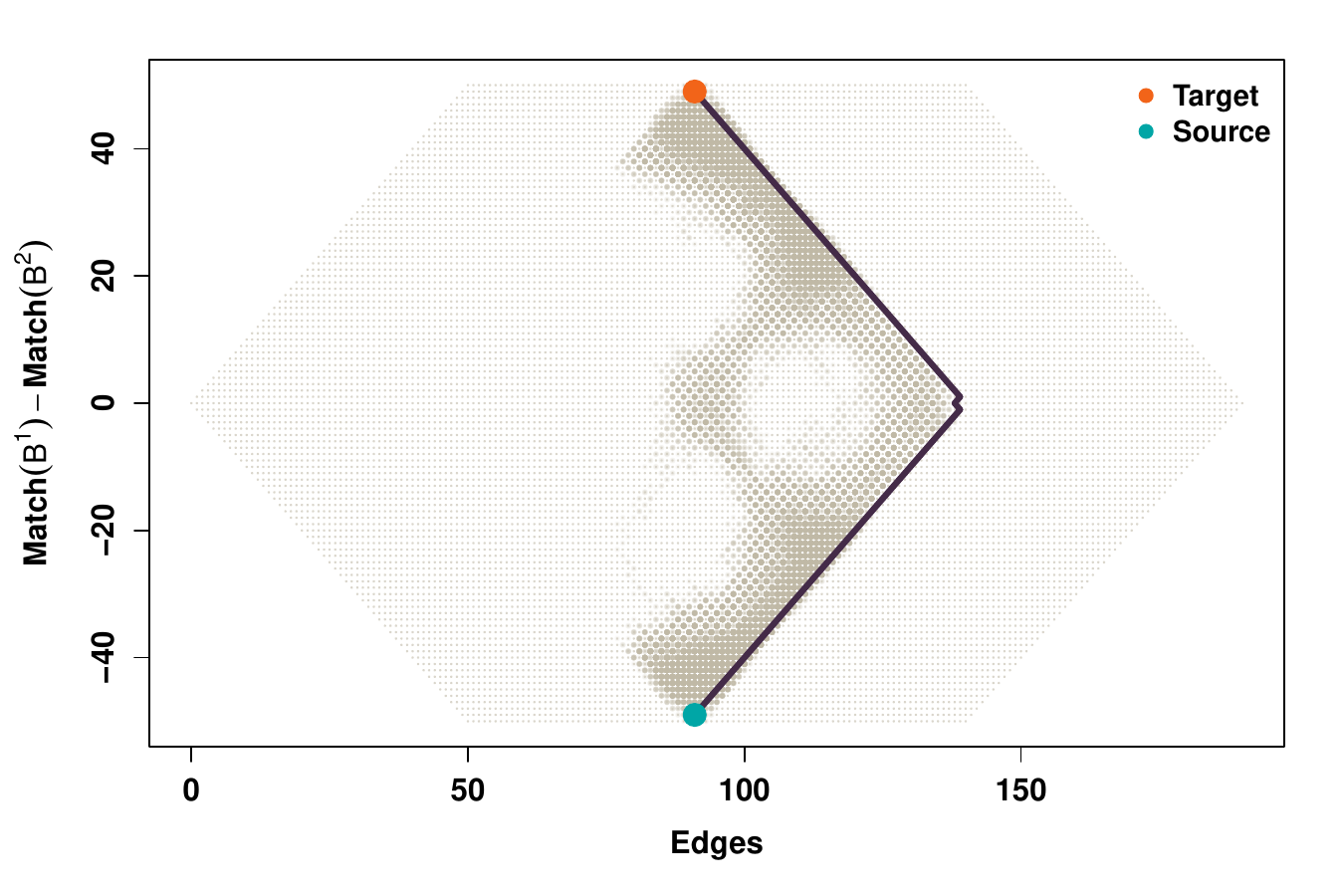}
\caption{State space of the faction alignment model, projected onto dimensions of edge count and polarization. Darker shaded areas indicate regions of higher state probability.  Purple line indicates MSPCP between source and target states. \label{f:demo_ss_mspcp}}
\end{figure}

To apply the theory of Sec.~\ref{sec:theory} to our case, we begin by defining our state space, and estimating the associated state probabilities.  $\mathcal{Y}$ is here the set of all labeled 20-node graphs with our specified covariate assignments, and we take $S$ to be the set of all unique vectors of sufficient statistics for the ERGM of Eq.\ref{eq:model}.  We begin by taking a sample of 100 overdispersed seed graphs from the model; we do this by simulating a MCMC trajectory (burn-in $10^5$, thinning interval $10^5$) from a ``heated'' model with parameter vector $\theta_h = \theta/10$.  This flattens the graph potential (while retaining its overall shape), ensuring that the Markov chain can mix well over the entire graph space (verified by inspection of sample statistics).  From each of these seed graphs, we then take an MCMC trajectory of length $10^7$ (for a total of $10^9$ observations), recording the state at each transition.  (These trajectories are not thinned, as we wish to observe the Hamming connectivity among states.)  We estimate $S$ by the set of all unique observed states ($|\hat{S}|=164,462$), and $T_D$ by the set of all observed transitions.  Since Hamming moves (edge additions/removals) are reversible, all observed transitions are treated as symmetric (i.e., we know that $s\to s'$ and $s' \to s$ are both allowable if either is observed).  On average, each observed state has allowable transitions to approximately 6.7 other states in $S$.  Finally, we estimate state probabilities by frequency of appearance in the final MCMC sample.  State probabilities (projected onto a subset of statistics) are shown in Fig.~\ref{f:demo_ss_mspcp}.  All modeling and related calculations were performed using the \texttt{ergm} \citep{krivitsky.et.al:jss:2023,hunter.et.al:jss:2008}, \texttt{sna} \citep{butts:jss:2008b}, and \texttt{network} \citep{butts:jss:2008a} packages from the \texttt{statnet} \citep{handcock.et.al:jss:2008} library for the \textsf{R} statistical computing system \citep{rteam:sw:2026}.

To obtain the maximum state probability change path between the two alignments, we begin by selecting the highest probability state in each regime; candidate states were chosen by taking states for which the respective nodematch statistics for the two bases were above 80 (within-group) and below 50 (cross-group).  (These thresholds were determined by examining the distribution of model statistics.  Note that, as shown in Fig.~\ref{f:demo_ss_mspcp}, the most probable states are far from the boundary of possible states.)  Representative graphs for each high-probability state are shown in Fig.~\ref{f:demo_aligned}.  Given the two states, the MSPSP is then obtained using Dijkstra's algorithm (see Fig,~\ref{f:demo_ss_mspcp}).  Properties of states encountered along the path are subsequently tabulated, allowing us to examine the MSPCP-predicted change process.

\subsubsection{The MSPCP Follows the ``High Road''}

The shaded region of Fig.~\ref{f:demo_ss_mspcp} shows the state probabilities for the affiliation model, projected onto two dimensions: the edge count, and the difference between the two nodematch statistics.  The former obviously expresses overall affiliation activity, while the latter indicates the degree of polarization around the two respective bases of solidarity.  As noted above, one might envision realignment occurring through a ``high (density) road,'' in which between-group ties are added first and then ties to former allies or lost, or a ``low (density) road,'' in which solidarity is first eroded within-group, before ties are added to the opposite faction.  These respective scenarios correspond to change paths following the right versus the left side of Fig.~\ref{f:demo_ss_mspcp}.  An examination of the state probability distribution for the affiliation model immediately suggests that the low road is implausible: such a path requires the network to traverse a very large region of very low probability.  By contrast, there are numerous well-connected regions of high probability on the high-density side of Fig~\ref{f:demo_ss_mspcp}, suggesting much more favorable terrain for the high road.

Indeed, this turns out to be the case.  With the two highest-probability polarized states indicated in teal and orange (respectively), we can see that the MSPCP (purple line) connecting these states follows the high road.  It is evident from the plot that density is initially added by exclusively creating cross-group ties until a maximum density is reached, at which point within-group ties are removed until the network arrives at the opposite state.  The change path theory thus suggests that faction realignment is driven by an increase in solidarity across factional lines, rather than erosion of within-faction solidarity.

\subsubsection{Several Intermediates and Transition States Lie Along the MSPCP}

\begin{figure}
\centering
\includegraphics[width=0.7\textwidth]{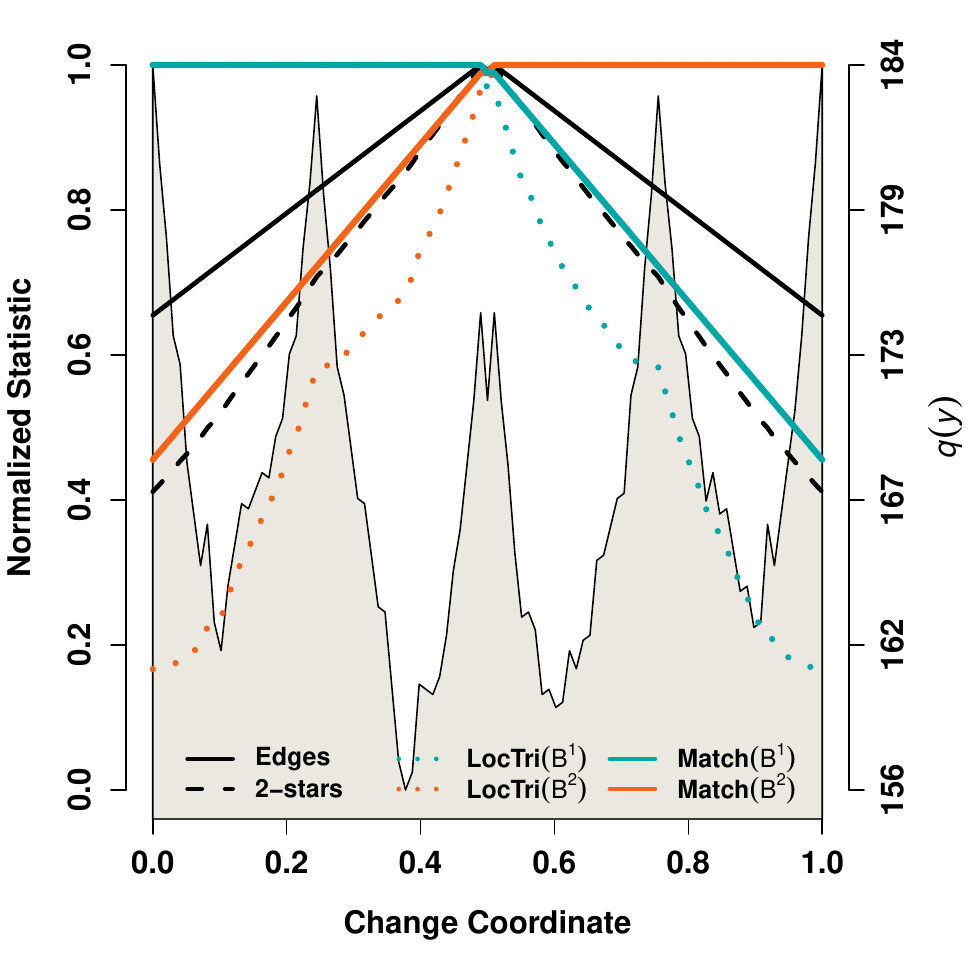}
\caption{Graph potential (shaded region/right axis) and normalized graph statistics (left axis) for the faction model, along the MSPCP. \label{f:demo_stats_mspcp}}
\end{figure}

We can gain further insight by examining the states traversed along the change path.  Fig.~\ref{f:demo_stats_mspcp} shows various graph statistics along the MSPCP, indexed by the change coordinate; the left axis shows normalized statistics (divided by their maximum values along the path), while the right axis shows the graph potential ($q$, indicated by shaded region).  We immediately note several intermediate states at roughly 25\%, 50\%, and 75\% of the way along the path, each of which is bounded by deep wells (transition states).  Of these, the outermost intermediates are only slightly less favorable than the endpoint states, while the central intermediate is considerably less favorable; the barriers that must be crossed to enter the central intermediate state are also higher than the barriers to enter the outer intermediate states.  We would thus anticipate that changes will occur in discrete stages, with transitions between the three intermediates occurring quickly and relatively long dwell times at the respective intermediate states.  Because of the difficulty of crossing the transition states to the middle intermediate, we would also predict that the change process would be particularly likely to fail at the first intermediate, with the system relaxing back to the previous aligned state rather than advancing to the middle, neutral state.  Once the system crosses to the third intermediate, however, the chance of a return prior to completion is expected to be very small.

This qualitative pattern can be further illuminated by examining the graph statistics along the change path (Fig.~\ref{f:demo_stats_mspcp}).  Moving from left to right, we see that the initial transition involves adding cross-faction members who are matched on $B^2$, while preserving affiliations matched on $B^1$.  The cross-group ties are non-random, as can be seen from the sharp upward slope in $B^2$ local triangulation: they rather involve formation and enlargement of \emph{cliques} that cross the faction boundary. In particular, inspection of graph draws reveals that solidarity is first added among one of the cross-group ``sides'' of $B^2$, with the symmetry broken in practice by whichever happens to gain greater density first.  Thus, while the addition of cross-faction ties increases both the edge count and the 2-star count (which are unfavorable), it becomes increasingly compensated for by within-attribute triangle formation.  This trade-off reaches an optimum at the first intermediate state, at which point one of the two attribute groups is fully saturated, and additional cross-faction ties must ``start over'' vis a vis formation of within-group coalitions; meanwhile, the cost of edges and 2-stars continue to mount.  As ties among the second group accumulate and the returns to triangle count increase again, the potential increases, until we arrive at the middle transition. At this point, solidarity along $B^1$ and $B^2$ is equal, but the high number of affiliations carried by each individual makes the state relatively unfavorable.  The next phase in the transition thus involves shedding affiliations, but this time along $B^1$.  This is at first unfavorable due to the loss of $B^1$ triangulation, but begins to become more favorable as degree costs decline.  This trade-off culminates at the third intermediate state, where we have maximum solidarity along $B^2$ combined with some cross-faction cliques among those aligned on $B^1$.  While the system may dwell in this state for an extended period, further breaking of cross-faction triangles makes many of the remaining ties unfavorable, driving the system to the destination state.  This state, a mirror of the source state, has factions that are solidly split along the $B^2$ axis, with a relatively small number of cross-faction ties among those aligned on $B^1$.  (This process is graphically summarized below in Fig.~\ref{f:unified}.)

\subsection{Comparison with Models of Explicit Dynamics} \label{sec:egp_compare}

Applying the MSPCP to the faction realignment model paints a vivid picture of how realignment might occur, based on fairly minimal information (the cross-sectional network distribution, combined with the Hamming connectivity of states).  How well, however, does this approximate the behavior of explicit models of network dynamics?  Here, we consider four such models, comparing the transition behavior of each to what would be predicted by the MSPCP.  As our model is specified in ERGM form, we employ EGPs (which ensure that their local dynamics are consistent with the long-run graph distribution); per Sec.~\ref{sec:theory}, we consider dynamic processes that tend to follow the potential surface, and for which the MSPCP could be expected to provide a reasonable approximation.  Specifically, the processes we use here may be briefly summarized as follows:

\begin{description}
\item[Longitudinal ERGM (LERGM)] $Y$ transitions between Hamming-adjacent states with a rate function that is sigmoidal in the difference between state potentials.  \citet{koskinen.lomi:jsp:2013} observe that the LERGM can be interpreted as a SAOM-like stochastic choice process in which opportunities to update relationships arise uniformly at random, with $q(y')-q(y)$ being the difference in utility for the tie-changing actor(s) when moving from graph $y$ to graph $y'$.
\item[Change Inhibition process (CI)] $Y$ transitions between Hamming adjacent states, with $q$-increasing moves occurring at a constant rate and $q$-decreasing moves occurring at a rate that falls exponentially with the potential difference.  The CI process can be motivated by a boundedly rational behavioral model in which relationships are randomly perturbed by exogenous events, with individuals motivated to repair or inhibit changes to the extent that they result in utility loss (with, as above, $q(y')-q(y)$ being the difference in utility for the tie-changing actor(s) when moving from graph $y$ to graph $y'$).
\item[Constant Dissolution Continuum STERGM (CDCSTERGM)] Edges are formed within $Y$ at a rate based on the resulting change in $q$, while being lost at a constant rate.  Note that this process tends to move uphill when adding edges, but is insensitive to the potential surface with respect to edge decay.  This family is thus an interesting ``borderline'' case for the likely applicability of the MSPCP.
\item[Constant Formation Continuum STERGM (CFCSTERGM)] Edges are \emph{lost} within $Y$ at a rate based on the resulting change in $q$, while being \emph{added} at a constant rate.  The counterpart of the CDCSTERGM, this process is sensitive to $q$ for tie decay, but not for edge addition.  It is thus unclear \emph{ex ante} whether the MSPCP will accurately predict its change paths.
\end{description}

All four models are detailed in \citet{butts:jms:2024}.  For the CDCSTERGM and CFCSTERGM, we set the respective dissolution/formation parameters to 0.5, adjusting the edge parameter of the formation/dissolution model accordingly to maintain the target equilibrium behavior.  Although these models differ considerably vis a vis their detailed dynamics, we shall see that they ultimately achieve faction realignment in very similar ways.

To examine the transition behavior of each model, we begin by drawing a random sample of 50 graphs from the maximum probability $B^1$-aligned state.  This is done by rejection sampling: we sample 500 graphs from an MCMC trajectory with a burn-in of $1.6 \times 10^5$ iterations and a thinning-parameter of $10^4$ iterations, keeping draws that are in the target state.  We extend this trajectory by an equivalent number of draws (again, keeping those in the target state) until 50 graphs are accumulated.  For each sampled graph, we then simulate forward trajectories using each of the four EGPs.  Trajectories returning to the initial state are truncated and simulated forward, with the process being continued until the trajectory reaches the target ($B^2$-aligned) state.  The resulting trajectories, by construction, originate in the source state and terminate in the target state.  We retain the 50 trajectories from each seed graph for each EGP for further analysis.  (Simulation was performed using the \texttt{ergmgp} package \citep{butts:sw:2023}.)

Note that, while every trajectory extends from the origin to the destination state, the trajectories are \emph{walks} rather than paths: trajectories may backtrack and/or pursue transient excursions before returning to the main source/target path.  To compare the underlying path followed during the change process, we thus prune off-pathway behavior from the trajectories by expressing the trajectory as a directed graph on $S$ and extracting the source/target geodesic.  

\subsubsection{The MSPCP Describes the Most Common Path for All Models}

\begin{figure}
\centering
\includegraphics[width=\textwidth]{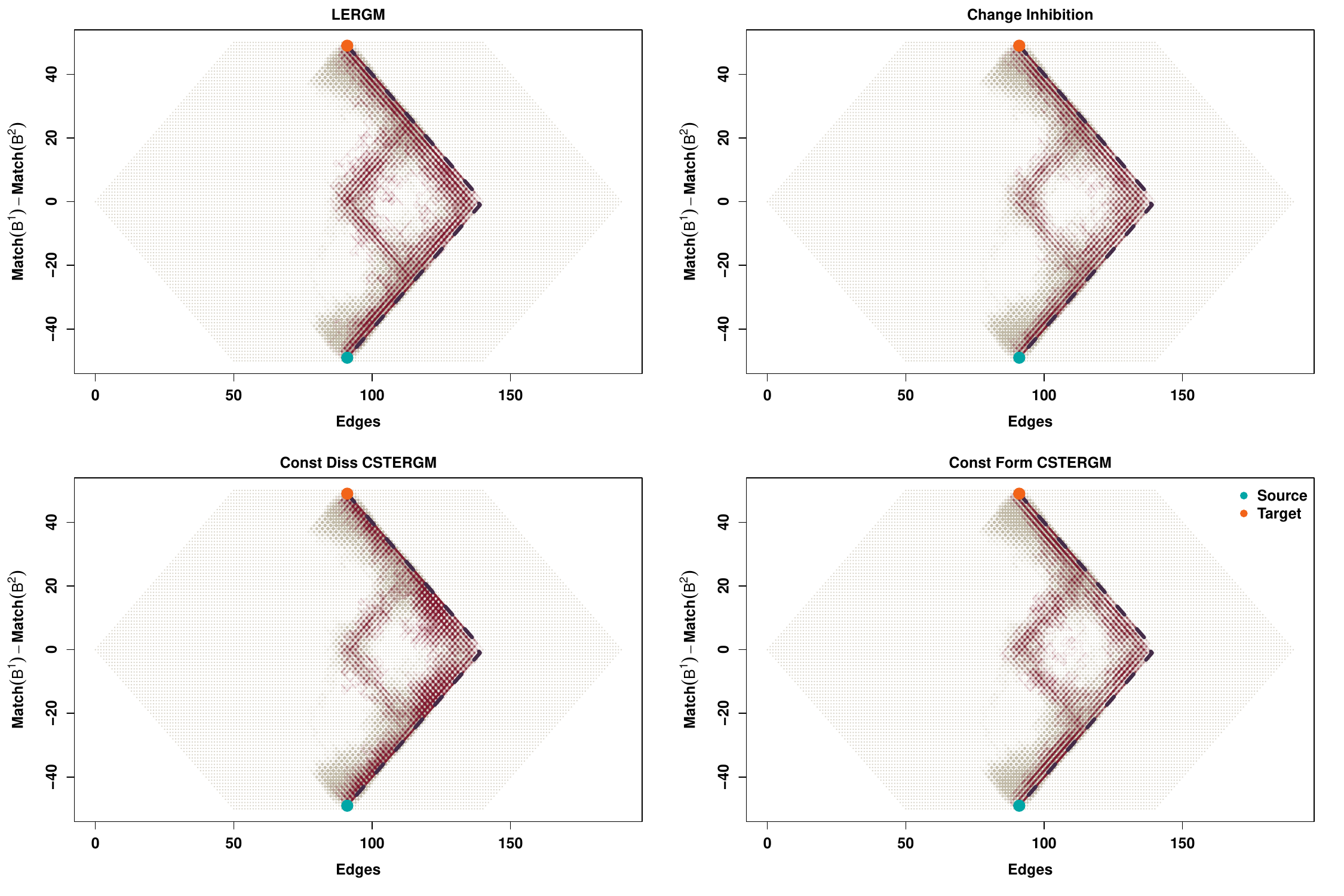}
\caption{Distribution of change paths (red) by dynamic process for the faction alignment model.  Dotted purple line shows the MSPCP.  The majority of change paths lie close to the MSPCP, with a fraction following a secondary, ``M-shaped'' change path. \label{f:demo_traj}}
\end{figure}

Fig.~\ref{f:demo_traj} shows the change path distribution for each of the four EGPs; darker-shaded segments appear more frequently over the 50 sample trajectories.  As predicted by the MSPCP, all observed trajectories follow the ``high road:'' faction realignment always begins with the formation of cross-faction ties, and never with the loss of in-group solidarity.  Moreover, as can be seen, the majority of paths hew closely to the MSPCP: overall, approximately 75\% of all trajectories lie in the ``bundle'' of MSPCP-adjacent paths which involve edge addition until ingroup and outgroup degrees are equalized, followed by deletion of former in-group ties.  Despite their substantial differences in underlying dynamics, the MSPCP is consistently able to discover the primary mode of faction realignment for all four models.

\begin{figure}
\centering
\includegraphics[width=\textwidth]{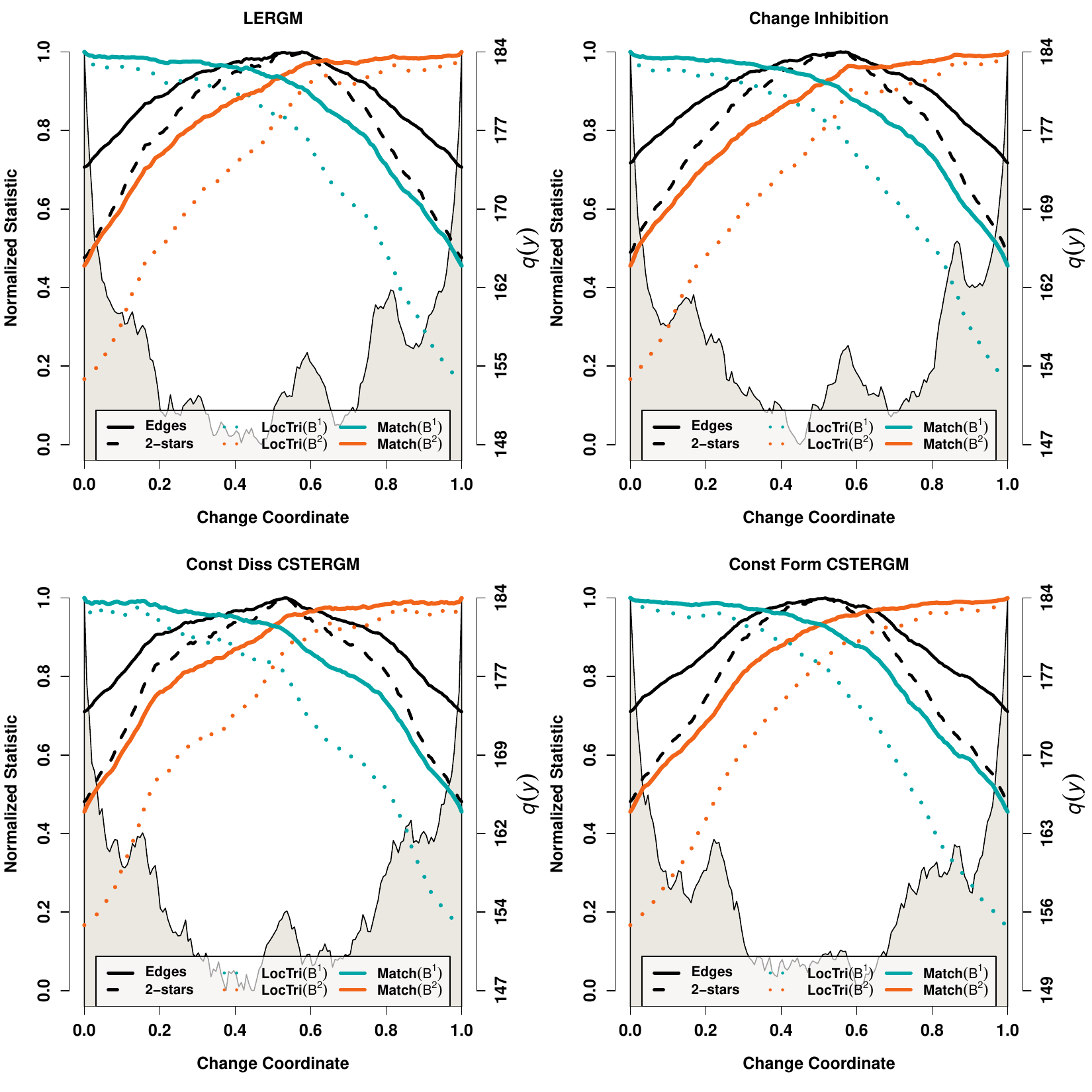}
\caption{Mean graph potential (shaded region/right axis) and normalized graph statistics (left axis) for aligned trajectories along the primary change path, by EGP. Although ``softer,'' all trajectories show similar patterns of behavior to the MSPCP.  \label{f:demo_traj_egp_mspcp}}
\end{figure}

Given that the MSPCP is an ideal that is not exactly followed, it is useful to compare the mean statistics along the change paths taken by the EGP trajectories with those of the MSPCP.  As the simulated paths vary somewhat in length (leading some to hit the same landmarks slightly earlier or later), we first align them along the change coordinate by expanding or contracting the change coordinate ``time'' between successive states (versus equal division) in each trajectory so as to minimize the mean RMSE of the graph potential between the aligned trajectories.  (States between observed steps were interpolated by splines.)  The resulting means for the MSPCP-adjacent trajectories are shown in Fig.~\ref{f:demo_traj_egp_mspcp}.  While the EGP statistics are somewhat softened relative to the MSPCP by regression to the mean, all EGPs show qualitatively similar behavior to the idealized path.  We thus see that the MSPCP is able to provide a general picture not only of the general mode of realignment, but also of the particular changes that occur.

\subsubsection{A Single Secondary Change Path Accounts for Residual Behavior}

It can be seen from Fig.~\ref{f:demo_traj} that, while all observed trajectories follow the high road, a minority (apx 25\%) of them deviate substantively from the MSPCP.  These trajectories follow a secondary change path, which coincides with the MSPCP up to/after the two outer intermediates, but that deviates in the transition between these intermediates.  In particular, the secondary change path achieves equalization after the initial period of cross-faction tie formation by a wave of within-faction erosion.  This leads to an alternative intermediate whose density is close to the source and target states, but that is faction-equalized.  From this alternative central intermediate, realignment then proceeds via another wave of tie addition (taking it to the second outer intermediate), with a final removal of cross-faction ties completing the change.  

\begin{figure}
\centering
\includegraphics[width=\textwidth]{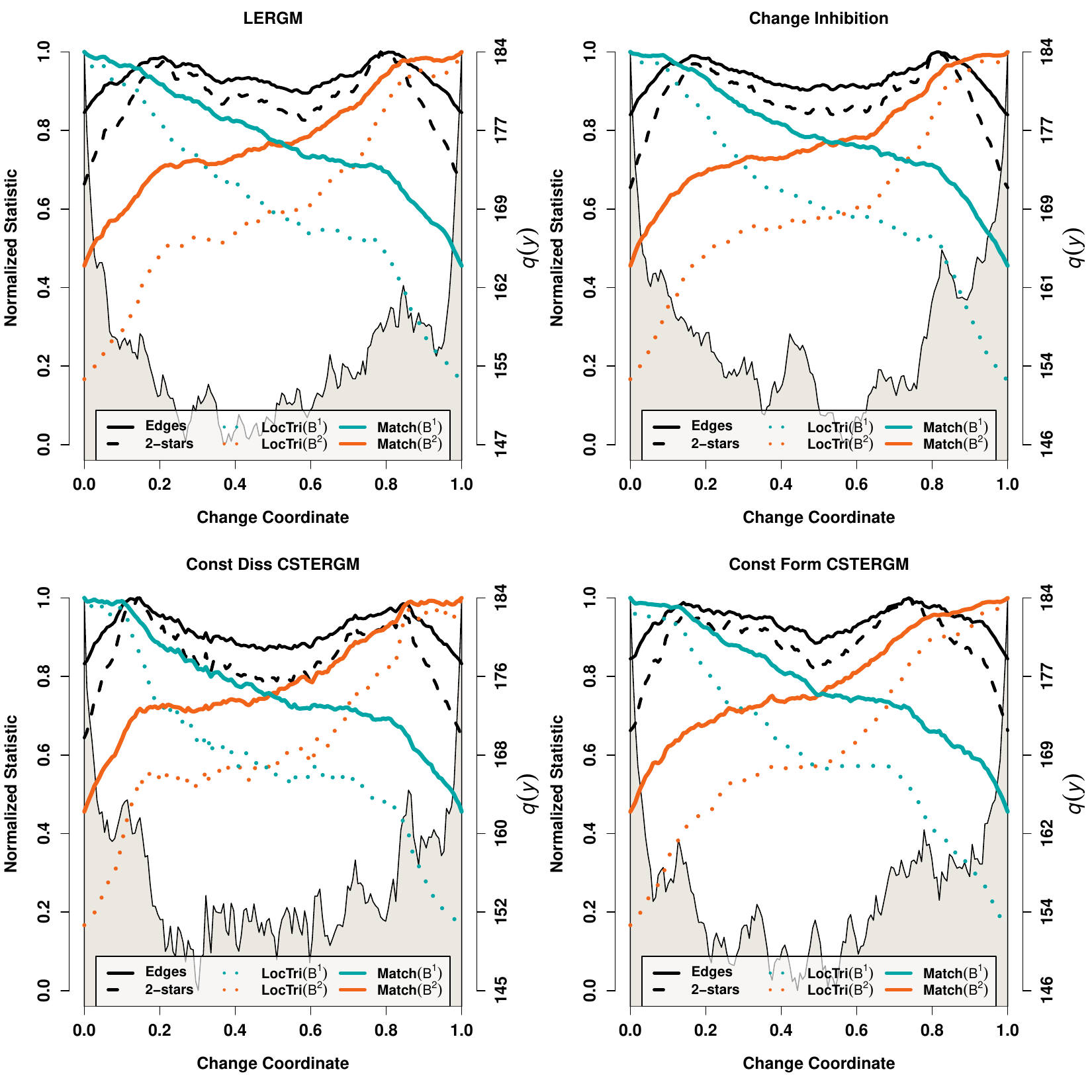}
\caption{Mean graph potential (shaded region/right axis) and normalized graph statistics (left acis) for aligned change paths from the secondary change path.  The ``M-shaped'' pattern of edge addition and removal is clearly visible. \label{f:demo_traj_egp_sec}}
\end{figure}

Additional detail can be had by examining mean graph statistics along the secondary change path, as shown for each EGP in Fig.~\ref{f:demo_traj_egp_sec}.  As before, trajectories following the secondary path were first aligned to minimize mean between-trajectory RMSE in graph potential, and with means of statistics along the change coordinate then taken for the aligned trajectories.  As expected, the patterns seen are very similar to those of Fig.~\ref{f:demo_traj_egp_mspcp} (and to the MSPCP) for the first and last phases of evolution, while deviating between the two outermost intermediates.  The ``M-shaped'' pattern of alliance formation and erosion is clearly evident, as is the overall lower potential of the states through which the path travels.  There is also a certain degree of irregularity in the mean potential of the states traversed in the middle portion of the change path, though the mean graph statistics remain fairly consistent; this suggests, near the middle intermediate, the presence of a plateau of relatively similar states through which the trajectories move irregularly.  

Inspection of Fig.~\ref{f:demo_traj} further reveals that the secondary path follows a second ``ridge'' of high state probability, as would be expected from the same principles that motivate the MSPCP.  It is thus reasonable to suspect that such paths may be discoverable \emph{ex ante,} using similar techniques.  We return to this issue in Sec.~\ref{sec:secondary}.

\subsubsection{Model Choice Affects Transient Behavior}

\begin{figure}
\centering
\includegraphics[width=0.5\textwidth]{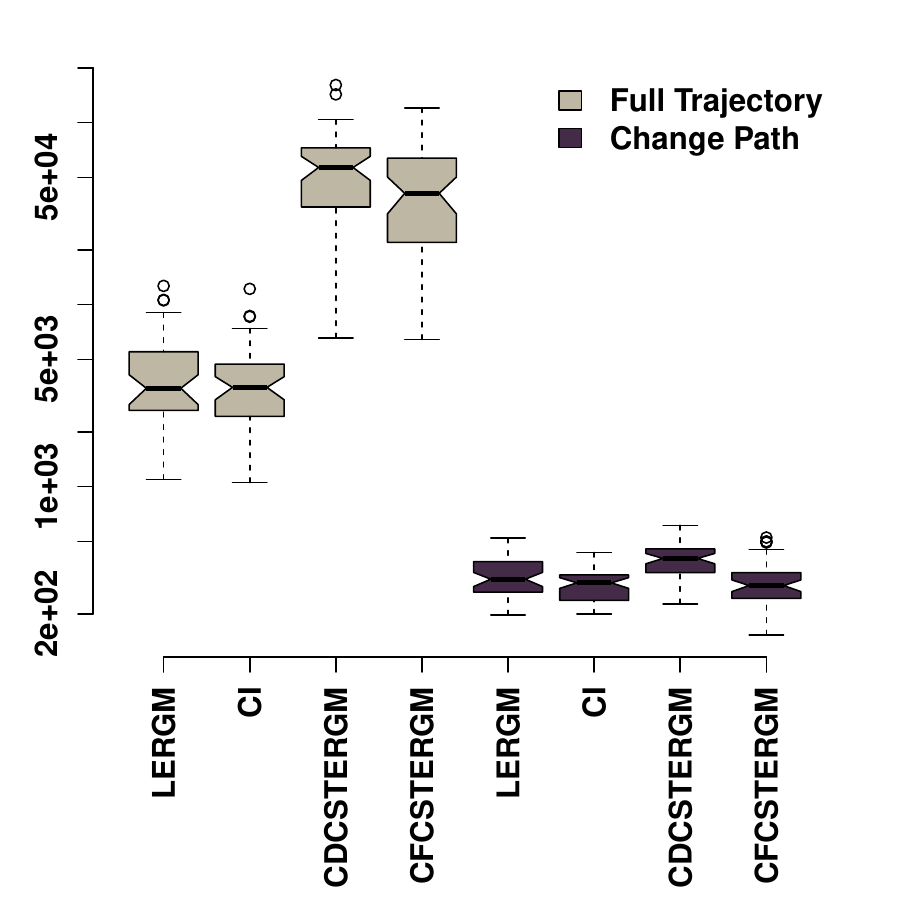}
\caption{Distribution of trajectory and change path lengths for the faction alignment model, by generative process. Light-colored boxplots (left) show lengths for the entire source/target walk, while purple boxplots (right) show lengths for change paths after pruning cycles and transients.  Non-overlapping notches indicate significant differences in medians by the procedure of \citet{mcgill.et.al:tas:1978}. \label{f:demo_traj_length}}
\end{figure}

Although all models considered here follow very similar change paths, this does not imply that their dynamics are the same.  Most notably, they differ slightly in their rate of adherence to the MSPCP, with the CDCSTERGM having the highest rate (84\%) followed by the CI (74\%), the CFSTERGM (72\%), and finally the LERGM (66\%).  Path and trajectory lengths also vary considerably.  Fig.~\ref{f:demo_traj_length} shows the distributions of these respective lengths for each dynamic process.  While the underlying paths are of roughly similar average length (varying from 292 for the CI process to 405 for the CDCSTERGM), total trajectory lengths vary by an order of magnitude, with the CSTERGM processes having roughly 100 times the lengths of their underlying paths, versus a factor of roughly 10 for the LERGM and CI processes.  The former thus ``wander'' considerably en route to the target, relative to the latter, possibly owing to the fact that certain moves (respectively, edge deletions and edge additions) occur at a constant rate that is insensitive to the potential landscape.  These trajectories are regularly forced to take unfavorable moves (and then find their way back), which is much less true of the LERGM and CI processes.  The fact that the qualitative behavior of such diverse processes can be accounted for in a common way is encouraging vis a vis the robustness of the approach.

\section{Discussion} \label{sec:discussion}

\subsection{The Steps to Faction Realignment}

\begin{figure}
\centering
\includegraphics[width=0.7\textwidth]{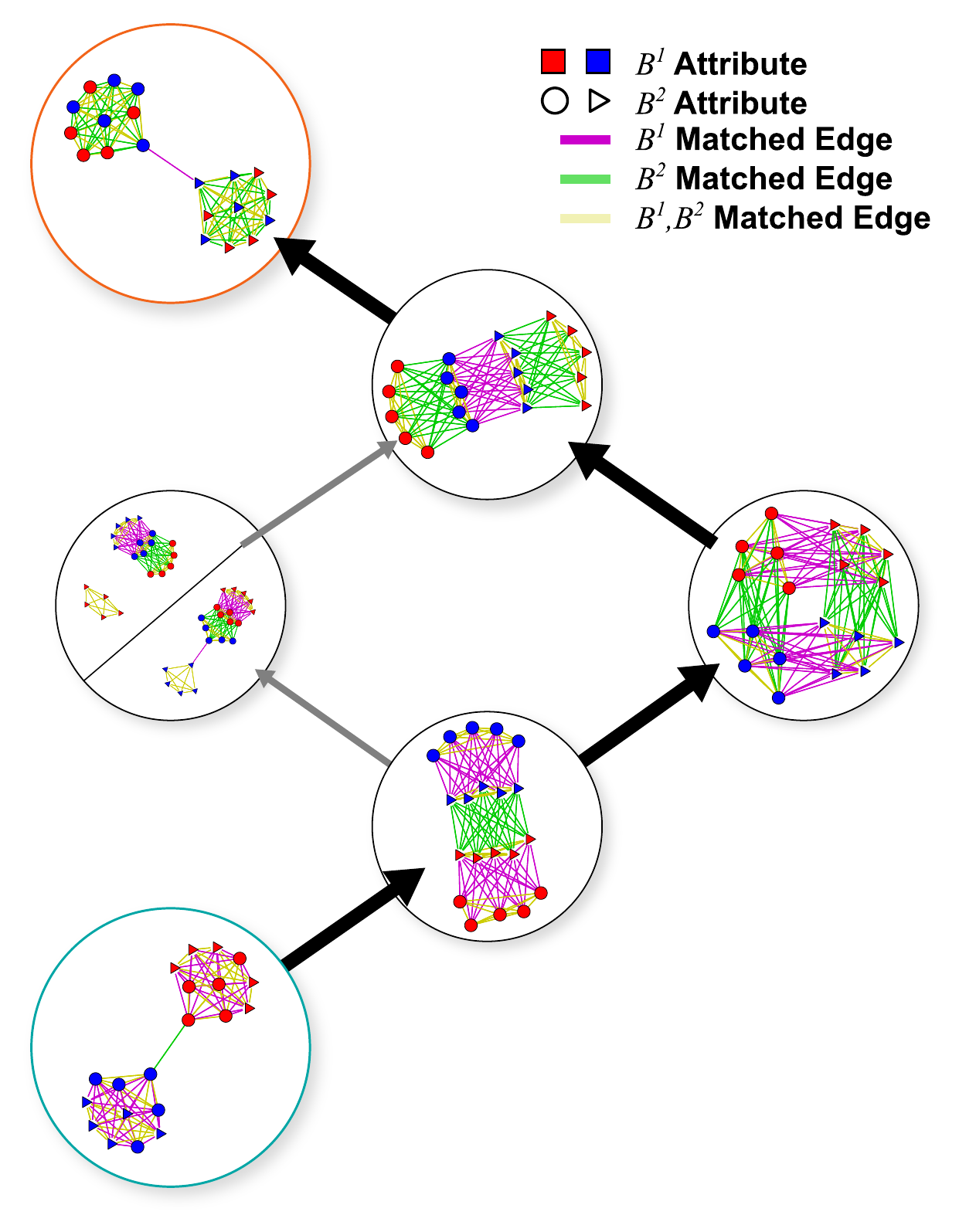}
\caption{Qualitative picture of faction realignment under the faction model. Graphs depict typical realizations at the source, target, and intermediate states. \label{f:unified}}
\end{figure}

Putting together the pieces of our analysis from Sec.~\ref{sec:demo}, Fig.~\ref{f:unified} provides a qualitative summary of the process of realignment under the faction model.  From an initially $B^1$ aligned state (with few cross-faction ties), cross-faction ties begin to be added among individuals sharing one of the two $B^2$ attributes; consolidation of solidarity within this latter group is reached at the first intermediate.  Progress to the second intermediate is made by the formation of cross-faction ties among those with the other $B^2$ attribute, leading to a ``four cycle of cliques'' bridging those matching on each attribute.  Loss of solidarity among members of one of the two $B^1$ groups leads to the third intermediate, with the target state reached when nearly all non-$B^2$ matching alliances are broken.  While, as we have seen, more realistic models of explicit dynamics lead to somewhat ``noisier'' behavior, they broadly follow this basic pattern.   We also note the presence of a second pattern that uses an alternative central intermediate; while this intermediate is less trivial to characterize, it appears to involve partial fragmentation of the system into a mixed component and a homogeneous component.  The fragmented state is ultimately less favorable to the symmetrically aligned state, leading to progress along the change path.

\subsection{Finding Secondary Change Paths} \label{sec:secondary}

While the maximum state probability change path was seen in this case to account for the majority of trajectories, a significant minority of trajectories followed a second change path following a lower ``ridge'' of relatively high-probability states.  Can such secondary paths be predicted \emph{ex ante}?  Although we do not pursue this here, we suggest that heuristic search using e.g. annealing methods for paths that have high state probabilities but that are also far from the MSPCP may be effective at discovering candidates.  The central challenge is to (on the one hand) avoid simply enumerating paths that lie close to the MSPCP, while (on the other) also avoid enumerating paths of low state probability (which are unlikely to be used).  Finding the correct balance of penalties thus seems key.  At present, this remains a promising open problem for further work.

\section{Conclusion} \label{sec:conclusion}

Predicting the paths through which change may occur in social networks is a complex problem with no one solution.  Here, we suggest that - for some systems - it may be possible to obtain insights into network dynamics by combining information about graph distributions (obtained e.g. from cross-sectional models) with some simple assumptions about how networks are likely to evolve within the space of graphs.  This leads to a framework that is much like the well-known transition state theory of chemical kinetics (at least within its qualitative aspects).  The utility of the former in providing \emph{quantitative} insights regarding e.g. reaction rates suggests likewise the possibility that similar quantities (e.g., differences in log probability across transition states) could prove predictive in the network case.  This intriguing possibility remains to be investigated.  However, as we have shown simple ideas of this type can lead to \emph{qualitative} insights into the dynamics of network change, which are robust to the fine details of network microdynamics.  This would seem to be a powerful tool for studying changes in groups, organizations, or other settings for which cross-sectional network models are becoming widely used and available as tools for theoretical investigation.

\bibliography{ctb}

\begin{thebibliography}{}

\bibitem[Bearman et~al., 2004]{bearman.et.al:ajs:2004}
Bearman, P.~S., Moody, J., and Stovel, K. (2004).
\newblock Chains of affection: The structure of adolescent romantic and sexual
  networks.
\newblock {\em American Journal of Sociology}, 110:44--91.

\bibitem[Blau, 1970]{blau:asr:1970}
Blau, P.~M. (1970).
\newblock A formal theory of differentiation in organizations.
\newblock {\em American Journal of Sociology}, 35(2):201--218.

\bibitem[Butts, 2008a]{butts:jss:2008a}
Butts, C.~T. (2008a).
\newblock network: a package for managing relational data in {R}.
\newblock {\em Journal of Statistical Software}, 24(2).

\bibitem[Butts, 2008b]{butts:sm:2008}
Butts, C.~T. (2008b).
\newblock A relational event framework for social action.
\newblock {\em Sociological Methodology}, 38(1):155--200.

\bibitem[Butts, 2008c]{butts:jss:2008b}
Butts, C.~T. (2008c).
\newblock Social network analysis with sna.
\newblock {\em Journal of Statistical Software}, 24(6).

\bibitem[Butts, 2023]{butts:sw:2023}
Butts, C.~T. (2023).
\newblock {ergmgp}: Tools for modeling {ERGM} generating processes.
\newblock Electronic Data File.

\bibitem[Butts, 2024]{butts:jms:2024}
Butts, C.~T. (2024).
\newblock Continuous time graph processes with known {ERGM} equilibria:
  Contextual review, extensions, and synthesis.
\newblock {\em Journal of Mathematical Sociology}, 48(2):129--171.

\bibitem[Butts et~al., 2012]{butts.et.al:joss:2012}
Butts, C.~T., Acton, R.~M., and Marcum, C.~S. (2012).
\newblock Interorganizational collaboration in the {H}urricane {K}atrina
  response.
\newblock {\em Journal of Social Structure}, 13.

\bibitem[Butts and Carley, 2007]{butts.carley:jms:2007}
Butts, C.~T. and Carley, K.~M. (2007).
\newblock Structural change and homeostasis in organizations: A
  decision-theoretic approach.
\newblock {\em Journal of Mathematical Sociology}, 31(4):295--321.

\bibitem[Carley, 1991]{carley:asr:1991}
Carley, K.~M. (1991).
\newblock A theory of group stability.
\newblock {\em American Sociological Review}, 56(3):331--354.

\bibitem[Cohen, 1964]{cohen:bk:1964}
Cohen, J.~E. (1964).
\newblock {\em Casual Groups of Monkeys and Men: Stochastic Models of Elemental
  Social Systems}.
\newblock Harvard, Cambridge, MA.

\bibitem[Dill and Bromberg, 2010]{dill.bromberg:bk:2010}
Dill, K.~A. and Bromberg, S. (2010).
\newblock {\em Molecular Driving Forces: Statistical Thermodynamics in Biology,
  Chemistry, Physics, and Nanoscience}.
\newblock Garland Science, London, second edition.

\bibitem[Durkheim, 1984]{durkheim:bk:1893}
Durkheim, E. ([1893] 1984).
\newblock {\em The Division of Labor in Society}.
\newblock The Free Press, New York.

\bibitem[Gould, 2002]{gould:ajs:2002}
Gould, R.~V. (2002).
\newblock The origins of status hierarchies: A formal theory and empirical
  test.
\newblock {\em American Journal of Sociology}, 107(5):1143--1178.

\bibitem[Handcock et~al., 2008]{handcock.et.al:jss:2008}
Handcock, M.~S., Hunter, D.~R., Butts, C.~T., Goodreau, S.~M., and Morris, M.
  (2008).
\newblock {statnet}: Software tools for the representation, visualization,
  analysis and simulation of network data.
\newblock {\em Journal of Statistical Software}, 24(1):1--11.

\bibitem[Hanneke and Xing, 2007]{hanneke.xing:ch:2007}
Hanneke, S. and Xing, E.~P. (2007).
\newblock Discrete temporal models of social networks.
\newblock In Airoldi, E.~M., an~Stephen E.~Fienberg, D. M.~B., Goldenberg, A.,
  Xing, E.~P., and Zheng, A.~X., editors, {\em Statistical Network Analysis:
  Models, Issues, and New Directions: ICML 2006 Workshop on Statistical Network
  Analysis, Pittsburgh, PA, USA, June 29, 2006, Revised Selected Papers},
  volume 4503 of {\em Lecture Notes in Computer Science}, pages 115--125.
  Springer-Verlag.

\bibitem[Hunter et~al., 2008]{hunter.et.al:jss:2008}
Hunter, D.~R., Handcock, M.~S., Butts, C.~T., Goodreau, S.~M., and Morris, M.
  (2008).
\newblock ergm: A package to fit, simulate and diagnose exponential-family
  models for networks.
\newblock {\em Journal of Statistical Software}, 24(3).

\bibitem[Hunter et~al., 2012]{hunter.et.al:jcgs:2012}
Hunter, D.~R., Krivitsky, P.~N., and Schweinberger, M. (2012).
\newblock Computational statistical methods for social network analysis.
\newblock {\em Journal of Computational and Graphical Statistics}, 21:856--882.

\bibitem[Koskinen and Lomi, 2013]{koskinen.lomi:jsp:2013}
Koskinen, J. and Lomi, A. (2013).
\newblock The local structure of globalization: The network dynamics of foreign
  direct investments in the international electricity industry.
\newblock {\em Journal of Statistical Physics}, 151:523--548.

\bibitem[Koskinen and Snijders, 2007]{koskinen.snijders:jspi:2007}
Koskinen, J.~H. and Snijders, T.~A. (2007).
\newblock Bayesian inference for dynamic social network data.
\newblock {\em Journal of Statistical Planning and Inference}, 137(12):3930 --
  3938.
\newblock 5th St. Petersburg Workshop on Simulation, Part \{II\}.

\bibitem[Krackhardt, 1999]{krackhardt:rso:1999}
Krackhardt, D. (1999).
\newblock The ties that torture: Simmelian tie analysis in organizations.
\newblock {\em Research in the Sociology of Organizations}, 16:183--210.

\bibitem[Krivitsky and Handcock, 2014]{krivitsky.handcock:jrssB:2014}
Krivitsky, P.~N. and Handcock, M.~S. (2014).
\newblock A separable model for dynamic networks.
\newblock {\em Journal of the Royal Statistical Society, Series B},
  76(1):29--46.

\bibitem[Krivitsky et~al., 2023]{krivitsky.et.al:jss:2023}
Krivitsky, P.~N., Hunter, D.~R., Morris, M., and Klumb, C. (2023).
\newblock {ergm} 4: New features for analyzing exponential-family random graph
  models.
\newblock {\em Journal of Statistical Software}, 105(6):1--44.

\bibitem[Laidler and King, 1983]{laidler.king:jpc:1983}
Laidler, K.~J. and King, M.~C. (1983).
\newblock Development of transition-state theory.
\newblock {\em The Journal of Physical Chemistry}, 87(15):2657--2664.

\bibitem[Lusher et~al., 2012]{lusher.et.al:bk:2012}
Lusher, D., Koskinen, J., and Robins, G. (2012).
\newblock {\em Exponential Random Graph Models for Social Networks: Theory,
  Methods, and Applications}.
\newblock Cambridge University Press, Cambridge.

\bibitem[McGill et~al., 1978]{mcgill.et.al:tas:1978}
McGill, R., Tukey, J.~W., and Larsen, W.~A. (1978).
\newblock Variations of box plots.
\newblock {\em The American Statistician}, 32:12--16.

\bibitem[Pfeffer and Salancik, 1978]{pfeffer.salancik:bk:1978}
Pfeffer, J. and Salancik, G.~R. (1978).
\newblock {\em External Control of Organizations.}
\newblock Harper and Row, New York.

\bibitem[{R Core Team}, 2026]{rteam:sw:2026}
{R Core Team} (2026).
\newblock {\em R: A Language and Environment for Statistical Computing}.
\newblock R Foundation for Statistical Computing, Vienna, Austria.

\bibitem[Robins and Pattison, 2001]{robins.pattison:jms:2001}
Robins, G.~L. and Pattison, P.~E. (2001).
\newblock Random graph models for temporal processes in social networks.
\newblock {\em Journal of Mathematical Sociology}, 25:5--41.

\bibitem[Robins et~al., 2007]{robins.et.al:sn:2007}
Robins, G.~L., Pattison, P.~E., Kalish, Y., and Lusher, D. (2007).
\newblock An introduction to exponential random graph ($p^*$) models for social
  networks.
\newblock {\em Social Networks}, 29:173--191.

\bibitem[Schweinberger et~al., 2020]{schweinberger.et.al:ss:2020}
Schweinberger, M., Krivitsky, P.~N., Butts, C.~T., and Stewart, J. (2020).
\newblock Exponential-family models of random graphs: Inference in finite-,
  super-, and infinite-population scenarios.
\newblock {\em Statistical Science}, 35(4):627--662.

\bibitem[Simmel, 1898]{simmel:ajs:1898}
Simmel, G. (1898).
\newblock The persistence of social groups: {I}.
\newblock {\em American Journal of Sociology}, 3(5):662--698.

\bibitem[Snijders, 2001]{snijders:sm:2001}
Snijders, T. A.~B. (2001).
\newblock The statistical evaluation of social network dynamics.
\newblock {\em Sociological Methodology}, 31:361--395.

\bibitem[Stadtfeld et~al., 2017]{stadtfeld.et.al:sm:2017}
Stadtfeld, C., Hollway, J., and Block, P. (2017).
\newblock Dynamic network actor models: Investigating coordination ties through
  time.
\newblock {\em Sociological Methodology}, 47(1):1--40.

\bibitem[Tainter, 1988]{tainter:bk:1988}
Tainter, J.~A. (1988).
\newblock {\em The Collapse of Complex Societies}.
\newblock Cambridge University Press, Cambridge.

\bibitem[White, 1970]{white:bk:1970}
White, H.~C. (1970).
\newblock {\em Chains of Opportunity: System Models of Mobility in
  Organizations}.
\newblock Harvard University Press, Cambridge, MA.

\bibitem[Willer, 2007]{willer:jms:2007}
Willer, R. (2007).
\newblock The role of metanetworks in network evolution.
\newblock {\em The Journal of Mathematical Sociology}, 31(2):101--119.

\bibitem[Yu et~al., 2021]{yu.et.al:siam:2021}
Yu, Y., Grazioli, G., Phillips, N.~E., and Butts, C.~T. (2021).
\newblock Local graph stability in exponential family random graph models.
\newblock {\em SIAM Journal on Applied Mathematics}, 81(4):1389--1415.

\bibitem[Zachary, 1977]{zachary:jar:1977}
Zachary, W.~W. (1977).
\newblock An information flow model for conflict and fission in small groups.
\newblock {\em Journal of Anthropological Research}, 33(4):452--473.

\end{thebibliography}


\end{document}